\documentclass[aps,reprint,floatfix,pre,nofootinbib]{revtex4-2}

\usepackage[utf8]{inputenc}
\usepackage{graphicx}
\usepackage{amsmath,amssymb,amsfonts}
\usepackage{bm}

\usepackage{booktabs}
\usepackage{tabularx}

\usepackage{xcolor}
\usepackage{transparent}
\definecolor{linkColor}{RGB}{0,70,120}
\definecolor{darkgreen}{RGB}{0,128,0}
\definecolor{darkgray}{RGB}{90,90,90}

\definecolor{linkColor}{RGB}{0,100,150}
\usepackage[colorlinks=true, allcolors=linkColor,pdfborder={0 0 0},pdfencoding = auto]{hyperref}

\usepackage{listings}

\definecolor{bg}{gray}{0.95}
\definecolor{cbrule}{RGB}{255,179,102}    % = orange!70
\definecolor{cbkeyword}{HTML}{008000}
\definecolor{cbstring}{HTML}{BA2121}
\definecolor{cbcomment}{HTML}{3D7B7B}
\definecolor{cbbuiltin}{HTML}{0000FF}
\definecolor{cboutput}{HTML}{666666}

\lstdefinestyle{txcode}{
  language=Python,
  basicstyle=\ttfamily\small,
  keywordstyle=\bfseries\color{cbkeyword},
  stringstyle=\color{cbstring},
  commentstyle=\itshape\color{cbcomment},
  emph={jax,jnp,triangular,mesh,geometry,print},
  emphstyle=\color{cbbuiltin},
  moredelim=[l][\color{cboutput}]{>>>},
  numbers=left,
  numberstyle=\tiny\color{cboutput},
  numbersep=8pt,
  xleftmargin=25pt,
  framexleftmargin=25pt,
  backgroundcolor=\color{bg},
  frame=tb,
  framerule=2pt,
  rulecolor=\color{cbrule},
  framesep=3pt,
  aboveskip=8pt,
  belowskip=8pt,
  breaklines=true,
  breakatwhitespace=true,
  postbreak=\mbox{\textcolor{cbrule}{$\hookrightarrow$}\space},
  columns=fullflexible,
  keepspaces=true,
  showstringspaces=false,
  tabsize=4,
}

\lstnewenvironment{pythonbox}[1][]{%
  \setlength{\leftskip}{0pt}\setlength{\rightskip}{0pt}%
  \setlength{\parindent}{0pt}\setlength{\parfillskip}{0pt plus 1fil}%
  \lstset{style=txcode,#1}%
}{}

\usepackage[normalem]{ulem}

\usepackage{xspace}
\newcommand{\tx}{\texttt{triangulax}\xspace}
\newcommand{\jx}{\texttt{JAX}\xspace}
\newcommand{\memdg}{\texttt{Mem3DG}\xspace}

\begin{document}

\title{\tx: Differentiable Simulations with Triangular Meshes across Soft Matter Physics}

\author{Samuel Silliman}
\affiliation{Joseph Henry Laboratories of Physics, Princeton University, Princeton, New Jersey 08544, USA} 
\author{Qiwei Yu}
\affiliation{Lewis-Sigler Institute for Integrative Genomics, Princeton University, Princeton, NJ 08544}
\affiliation{Leinweber Institute for Theoretical Physics, Stanford University, Stanford, CA 94305}

\author{Nikolas Claussen}
\email{nc1333@princeton.edu}
\affiliation{Joseph Henry Laboratories of Physics, Princeton University, Princeton, New Jersey 08544, USA} 
\affiliation{Princeton Center for Theoretical Science, Princeton University, Princeton, New Jersey 08544, USA}

\begin{abstract}
    Many physical and biological systems --foams, membranes, elastic shells, or tissue sheets-- can be mathematically described as 2D surfaces. 
    These surfaces can host complex dynamics, like reaction-diffusion systems, and undergo drastic deformations.   
    Here, we present \href{https://github.com/nikolas-claussen/triangulax}{\tx}, an open-source Python library for simulations and geometry processing that discretizes surfaces using triangular meshes. 
    It combines two features.
    First, discrete differential geometry provides coordinate-free discretization, which allows simulating strongly deforming surfaces.
    Second, the library is built on the machine-learning framework \jx, so mesh-based energies and entire simulations can be differentiated automatically (including simulations with stochastic forces or topological modifications).
    This simplifies multi-physics simulations by computing forces automatically and allows fitting model parameters to experimental data or design objectives.
    We implement an algorithm to preserve mesh quality in strongly deforming surfaces and apply it to a model of membrane mechanics coupled to an on-membrane concentration field.
    We validate our approach against exact solutions and existing software, and demonstrate \tx across a range of problems, including membrane mechanics, vertex models, and inverse design of a shape-morphing elastomer sheet.
\end{abstract}

\maketitle

\section*{Introduction}

In soft matter and biological physics, many important systems are described by deformable two-dimensional (2d) surfaces, embedded in three-dimensional space. 
Examples include foam bubbles~\cite{Weaire.etal2005}, membranes~\cite{Seifert1997}, thin elastic shells~\cite{Efrati.etal2013}, and reaction-diffusion systems on curved surfaces~\cite{Turk1991}.
Triangular meshes are the most common way to represent 2d surfaces numerically and are ubiquitous in computer graphics and scientific computing.
In addition to the discretization of continuum models, triangular meshes also arise in inherently discrete models, for example, of cell tilings~\cite{Farhadifar.etal2007,Claussen.etal2026}. 
In these settings, surfaces can undergo drastic shape changes. Their dynamics are governed by complex, potentially non-equilibrium, ``multi-physics'' models.
A reusable and efficient simulation framework for triangular meshes that can handle large deformations is therefore highly desirable.
In addition, surface-based models are becoming increasingly complex, notably to accurately describe biological phenomena. For example, realistic models for pattern-formation on curved membranes easily require dozens of parameters~\cite{Brauns.etal2021}.
New computational tools are required to fit such models to experimental data, and more broadly explore possible behaviors across high-dimensional parameter space.
Methods from machine learning have shown promise in this task~\cite{Karniadakis.etal2021}.
 
Here, we present \href{https://github.com/nikolas-claussen/triangulax}{\tx}, a Python library for geometry processing and physical simulations with triangular meshes.
It is designed for modularity and reusability across simulation contexts. The library has two main features. 
First, \tx implements high-quality computational geometry tools developed in the field of \emph{discrete differential geometry} (DDG)~\cite{Crane.etal2013}. DDG builds on classical differential geometry to derive algorithms for geometry and mechanics on discrete meshes which are simple to implement and stable through large deformations. Second, \tx is built using \jx~\cite{jax2018github}, a numerical framework from the machine-learning community. This makes \tx compatible with just-in-time (JIT) compilation and \emph{automatic differentiation} (autodiff). 
JIT compilation delivers numerical performance in high-level Python, including by running code on GPUs.
Autodiff computes the gradient of any mesh-based function in an efficient and automated manner.
For example, autodiff can compute forces from arbitrary energies without tedious manual derivations, which makes \tx well-suited for multiphysics simulations like protein patterns on a deforming membrane~\cite{Arroyo.etal2018}.

Autodiff is more general and can evaluate the gradient of the outputs of complex simulations with respect to initial conditions or parameters like reaction rates or elastic moduli---the entire simulation is differentiable. This ability is a powerful tool for \emph{inverse problems}, where the parameters are optimized so the simulation output fits experimental data or fulfills a design objective.
Differentiable models have been used to design self-assembling colloids~\cite{Goodrich.etal2021}, and to explore self-organization in simple models of pattern formation~\cite{Mordvintsev.etal2020}.
There is growing interest in using autodiff to reverse-engineer principles of morphogenesis (biological shape-generation)~\cite{Deshpande.etal2025,Pahng.etal2025,Pasqui.etal2026}.
With \tx, we aim to bring this approach to simulations of tissue sheets~\cite{Farhadifar.etal2007}, membranes~\cite{Seifert1997}, and elastic shells~\cite{Efrati.etal2013}, all of which can be described as curved, 2d surfaces.

\subsection*{Related work}

The venerable \texttt{Surface Evolver}~\cite{Brakke1992} (last updated in 2008) allows users to minimize energies of surfaces described as triangular meshes by gradient descent. The \texttt{Surface Evolver} has proven highly useful to researchers in several disciplines. This success calls for a modern version with extended capabilities (like non-equilibrium dynamics).

Recently, differentiable simulators have been introduced to several areas of computational physics (e.g., molecular dynamics~\cite{Schoenholz.Cubuk2019} or finite elements~\cite{Xue.etal2023}) and 3D graphics~\cite{Jatavallabhula.etal2019}, allowing researchers to harness tools from machine learning for physical simulations.
Several high-quality modern geometry libraries exist (e.g., \texttt{CGAL}~\cite{cgal:eb-26a} or \texttt{libigl}~\cite{Jacobson.etal2018}), but do not feature autodiff (C++ codebases with limited Python bindings also impede rapid prototyping). This means gradients of energies or fitting loss functions must be derived and implemented by hand.
Furthermore, these libraries are not designed for physical simulations.
The \texttt{Mem3DG} software~\cite{Zhu.etal2022} is based on DDG, but also does not use autodiff. This makes it harder to extend it to free energy models.
Further, \texttt{Mem3DG} is specialized to membrane mechanics and does not cover elasticity or vertex models. We benchmark \tx against \texttt{Mem3DG} below and find improvements in speed and stability.

Purpose-built libraries, like the C++ soft-matter library SAMoS~\cite{Sknepnek2013,Barton.etal2017}, can be challenging to adapt to new use cases due to their custom infrastructure, and do not use autodiff. The recent \texttt{VertAX} library does leverage autodiff~\cite{Pasqui.etal2026}, but is specialized to equilibrium shapes of two-dimensional vertex models only.

\begin{figure*}[t]
    \centering
    \includegraphics[width=0.8\textwidth]{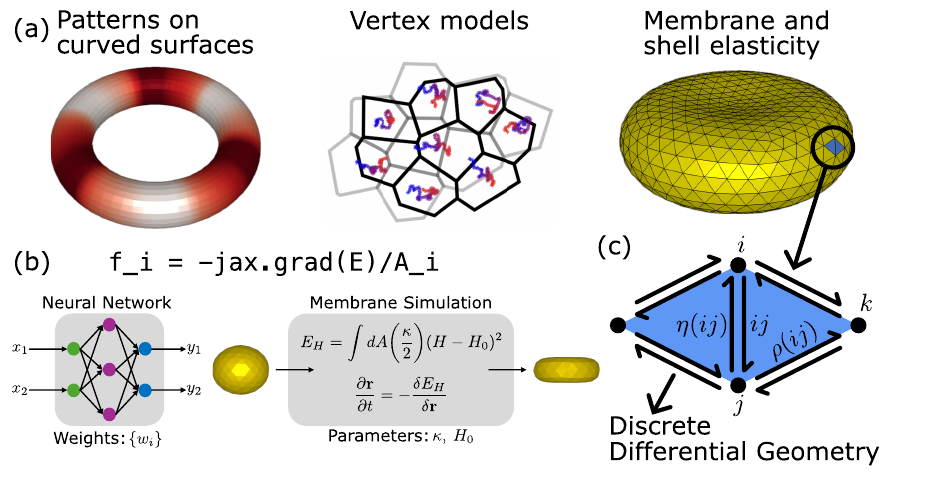}
    \caption{\textbf{\tx overview.}
    (a) Simulations of reaction-diffusion systems on surfaces, vertex models for tissue dynamics, and membrane mechanics all rely on triangular meshes. \tx provides a common simulation toolset for all of them.
    (b) \jx allows automatic differentiation of any function defined by computer code(\texttt{jax.grad}). This replaces manual calculation of the gradients of mesh-based energies, such as the force density arising from a free energy (\texttt{f\_i} is the force density at vertex \texttt{i} with area \texttt{A\_i}). Autodiff can also compute the gradients of entire simulations w.r.t. simulation parameters. This allows treating physical simulations like neural networks and optimizing their parameters to fit experimental data or fulfill design objectives. 
    (c) DDG translates differential geometry from the smooth to triangulated surfaces, and provides elegant, discrete versions of the smooth differential operators (e.g., Laplacian, gradient) and energies.
    To implement the DDG toolset, \tx represents surfaces as half-edge meshes with vertices $i, j$ and oriented edges $ij$. The \texttt{twin} $\eta(ij)$ and \texttt{next} $\rho(ij)$ operators specify how half-edges are connected.}
    \label{fig:abstract}
\end{figure*}

% error ~\pi/4.

\section*{Methods}
\renewcommand{\thesubsection}{\arabic{subsection}}

\subsection{A primer on discrete differential geometry}

Curved surfaces are traditionally studied as infinitely smooth manifolds, embedded by a map $\mathbf{r}$ into 2d or 3d Euclidean space.  Processing surfaces computationally, however, requires discretization. 
In physics, using triangulations to represent geometry in a coordinate-free way goes back to Regge's work on general relativity~\cite{Regge1961}.
Further, many physical systems feature discrete geometries (granular packings or cell tilings).
A discrete surface is a polygonal mesh. The mesh combinatorics are defined by a set of vertices $i, j$, edges $ij$, and triangular\footnote{Larger polygonal faces can always be divided into triangles.} faces $ijk$ (Fig.~\ref{fig:abstract}a,c), while the mesh geometry is defined by a set of vertex positions $\mathbf{r}_i$ (see Table~\ref{tab:notation_discrete} for notation details).

\begin{table}
    \centering
    \begin{tabular}{l|l}
    \toprule
    Symbol & Description \\
    \midrule
    $i$, $ij$, $ijk$ & vertices, edges, faces \\
    $\mathbf{r}_i$ & vertex positions \\
    $\mathbf{e}_{ij}=\mathbf{r}_j-\mathbf{r}_i$ & edge vectors \\
    $\mathbf{n}_{i}, \mathbf{n}_{ijk}$ & unit normals \\
    $a_i, a_{ijk}$ & vertex, face area \\
    $\alpha_{ij}, \beta_{ij}, \gamma_i$ & triangle angles \\
    $\varphi_{ij}$ & dihedral angles \\
    \bottomrule
    \end{tabular}
    \caption{Notation for triangulations.}
    \label{tab:notation_discrete}
\end{table}

DDG translates the mathematics of differential geometry into the setting of discrete surfaces. Developed with an eye to applications in computer graphics and video games~\cite{Crane.etal2013}, DDG provides discrete versions of differential-geometric objects (vector fields, metric tensors, or exterior forms) and operators (gradient, Laplacian, or shape operator). These can serve as building blocks for geometry processing and physical simulations, like discrete elastic energies~\cite{Grinspun.etal2003}. DDG also has deep connections to the physics of granular materials and cell tilings~\cite{Claussen.etal2026}.

% coordinate independent
Importantly, DDG is an \emph{intrinsic} framework that does not require a 2d coordinate chart.
By contrast, simulations on curved surfaces commonly choose a particular coordinate system (for example, the Monge parameterization uses a height function over the $xy$ plane), and discretize the resulting set of partial differential equations (PDEs) in the 2d coordinate plane using the finite-element method (e.g., Refs.~\cite{Winter.etal2025,Yu.Kosmrlj2025}).
This approach has two drawbacks:  strongly deforming surfaces can run into coordinate singularities (for example, overhangs in the case of a height function), complex topologies require multiple coordinate patches, and coordinate-based formulations can obscure the underlying geometry.
DDG overcomes these challenges by using intrinsic discretizations. 
For example, a covector field $\boldsymbol{\alpha}$ is discretized by its dot products with triangle edge vectors, $\boldsymbol{\alpha} \cdot \mathbf{e}_{ij}$, instead of its components $\alpha_1, \alpha_2$ in a coordinate chart.
DDG provides fast and stable discrete versions of quantities like the Gaussian curvature, which, in coordinates, require second-order derivatives.
SI Sect.~\ref{sec:DDG_SI} lists the most important DDG discretizations. SI Sect.~\ref{sec:convergence} and SI Fig.~\ref{fig:convergence} show their convergence under mesh refinement.

\subsection{Inverse problems using automatic differentiation}

Automatic differentiation is based on the chain rule for computing derivatives, $(g \circ f )' = (g'\circ f) \cdot f'$~\cite{Goodfellow.etal2016}.
An autodiff framework like \jx decomposes a piece of computer code into a sequence of elementary operations (e.g., addition of two variables, application of a trigonometric function), for each of which the derivative is implemented. The individual derivatives are then composed (using what is called ``forward-'' or ``reverse-mode'' autodiff) to compute the gradient of complex programs efficiently.
Autodiff can thus evaluate the gradient of the outputs of complex simulations with respect to initial conditions or parameters, like reaction rates or elastic moduli---the entire simulation is differentiable. This ability is a powerful tool for \emph{inverse problems}, where the parameters are optimized so the simulation output fits experimental data or fulfills a design objective.

Given a physical ``forward'' model with variables $\mathbf{x}$ and parameters $\boldsymbol{\theta}$, inverse problems ask to recover a set of parameters $\boldsymbol{\theta}$ that optimize an objective $\mathcal{L}$~\cite{Cao.etal2002}. The objective can be fitting experimental data or a design target, for instance, generating a surface with a specific shape.
More specifically, the forward model computes a simulation output $\mathbf{x}^*(\boldsymbol{\theta})$. 
For example, relaxing an energy $E$ with parameters such as elastic moduli leads to $\mathbf{x}^*(\boldsymbol{\theta}) = \arg\min_{\mathbf{x}}E(\mathbf{x};\boldsymbol{\theta})$. The inverse problem minimizes $\mathcal{L}[\mathbf{x}^*(\boldsymbol{\theta})]$ with respect to $\boldsymbol{\theta}$.

Autodiff allows writing \emph{differentiable simulators} in which the parameter gradient of the simulation output $\partial \mathbf{x}^*(\boldsymbol{\theta})/\partial \boldsymbol{\theta}$ can be computed automatically as the forward problem is solved~\cite{Newbury.etal2024}.
This enables gradient-based methods to find an optimal set of parameters, 
$\boldsymbol{\theta}^* = \arg\min_{\boldsymbol{\theta}}\mathcal{L}[\mathbf{x}^*(\boldsymbol{\theta})]$. Effectively, the forward simulation can be treated like a ``neural network'' which maps initial conditions to simulation results: parameters (like elastic moduli or reaction rates) are trainable like neural network ``weights''~\cite{kidger2021on}.
As we show below, autodiff also works for simulations that appear ``non-differentiable'' at first sight, including topological modifications and stochastic dynamics.

\subsection{The \tx library}

In \tx, the geometry (vertex positions) and connectivity (vertex adjacency) of triangular meshes are represented by separate \texttt{GeoMesh} and \texttt{HeMesh} classes. The \texttt{HeMesh} class represents mesh adjacency using the \emph{half-edge} data structure. In brief, a half-edge mesh stores a list of vertices, oriented edges, and faces. A set of tables encodes how the mesh elements are connected and allows navigation around the mesh (operators \texttt{next} and \texttt{twin} in Fig.~\ref{fig:abstract}c). Mesh connectivity can be modified through edge flips (T1 transitions), deletion, or splitting of vertices. These modifications can be used to remesh strongly deforming surfaces to improve numerical performance or implement events like cell rearrangement in discrete systems. \emph{Adjacency-based operators} enable neighborhood queries, like summing over all faces adjacent to a vertex. 

Combining adjacency information with mesh geometry allows computing \emph{geometric quantities} like normal vectors, edge lengths, Voronoi areas, and Gaussian curvature, which underlie DDG. \tx thus implements discrete versions of classic \emph{differential operators} like the exterior derivative, Laplace operator, or the area integral (the mass matrix). These operators can be applied to fields defined on mesh elements (for example, at each vertex). Our implementation follows~\cite{Meyer.etal2003}. Where possible, \tx uses intrinsic formulations that require only edge lengths $e_{ij}$. This approach makes \tx extensible; one can simulate dynamics in 2d (e.g., vertex models), 3d (e.g., membrane mechanics), or in periodic geometries all within the same framework (one only needs to specify a suitable distance function).

The entire \tx library is written for compatibility with the machine learning library \jx. This is ensured by using \jx-native control flow and a functional programming paradigm. Additionally, \tx meshes are stored as \jx \texttt{pytree} objects\footnote{\jx \texttt{pytree} objects are a general class of nested, container-like objects like lists of arrays.}. We provide efficient, vectorized implementations of geometric and differential qualities on triangular meshes. For example, neighborhood queries are carried out using a gather-scatter approach (see example in SI Fig.~\ref{sec:code_example}). This allows working with unstructured meshes in \jx arrays without resorting to workarounds like padding. \tx is compatible with both forward and reverse mode autodiff in \jx, as well as with JIT-compilation. It can automatically and efficiently compute the gradients of any mesh-based energy (like the ``premade'' energies provided in the \texttt{elastic} module). \tx seamlessly integrates with many \jx-based Python libraries. 
For example, we use the \texttt{optimistix} library~\cite{optimistix2024} for static simulations (energy minimization) and the \texttt{diffrax} library~\cite{kidger2021on} for dynamic ones (deterministic and stochastic differential equations). Both libraries implement automatic derivatives for their solvers, for example using the efficient adjoint method~\cite{Cao.etal2002}. This makes \tx simulations end-to-end differentiable.

\tx is hosted on GitHub and available as a PyPI package. Documentation and tutorials can be found on GitHub Pages. Table~\ref{tab:modules} lists the modules available in \tx, and SI Fig.~\ref{sec:code_example} shows a code example.

\subsection{Benchmarking}

We benchmarked \tx against \memdg, a package for solving mechanochemical dynamics of membranes using triangulated meshes~\cite{Zhu.etal2022}.
Using both packages to find the minimum-energy shape of deflated membrane vesicles (see Results Sect.~\ref{sec:membrane_elasticity}), we find that \tx achieves faster convergence to equilibrium shapes as the mesh resolution is increased, while maintaining similar or better mesh quality (SI Fig.~\ref{fig:benchmark supp} and SI Sect.~\ref{sec:benchmark}). Autodiff-based functionality was not compared since it is not provided by \memdg.

\section*{Results}
\setcounter{subsection}{0}

We now present case studies that apply \tx to systems across soft matter and biophysics. These tasks revolve around mesh-based energies, implemented using DDG. \tx automatically computes the gradients, making it easy to optimize or to compute forces.
Tutorials on membrane mechanics, vertex models, and gradient-based inverse design are available as Jupyter notebook tutorials \href{https://nikolas-claussen.github.io/triangulax/tutorials/00_triangular_meshes.html}{online}, where details on the simulated models and numerics can be found.

\subsection{Vertex models}

\begin{figure*}
    \centering
    \includegraphics{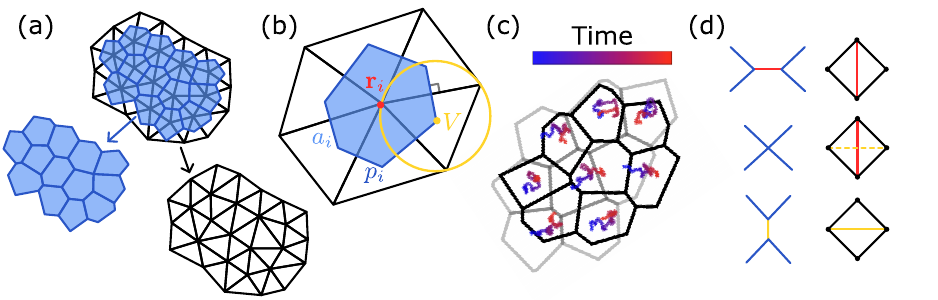}
    \caption{\textbf{Area-perimeter vertex models.}
    (a) A triangular mesh (black edges) and the dual Voronoi tessellation (blue polygons).
    (b) Zoomed-in view of a Voronoi cell with cell area $a_i$ and perimeter $p_i$. The Voronoi cells are centered at triangular mesh vertices $\mathbf{r}_i$ and have corners $V$ located at triangle circumcenters. Cell and triangle edges are orthogonal.
    (c) Trajectories of a cell group due to stochastic self-propulsion. 
    (d) When a cell-cell edge shrinks to length 0 (left series of images), a T1-event changes the mesh topology by flipping a triangulation edge (right series of images). Numerically, edge flip triggers can be computed from the triangulation angles~\cite{Barton.etal2017}.
    }
    \label{fig:vertex-models}
\end{figure*}

Vertex models are a class of biophysical models in which cells are represented as polygons or polyhedra, and thus parametrized by a set of vertices~\cite{Farhadifar.etal2007}. Vertex models are often applied to cells constrained to epithelial cell sheets, modeled as 2d polygonal tilings (Fig.~\ref{fig:vertex-models}a). These tilings can be represented by their \emph{dual graphs} in which cells form nodes, and adjacent cells are linked by edges.
This adjacency graph forms a triangulation (see Fig.~\ref{fig:vertex-models}a)\footnote{Higher-order faces can be triangulated by adding ``fictitious'' edges.}.

We used \tx to simulate the popular self-propelled Voronoi (SPV) vertex model~\cite{Bi.etal2016}. In the SPV, the cell tessellation is the Voronoi dual to a triangulation with vertices $\mathbf{r}_i$. The $\mathbf{r}_i$ are the cell centers. The Voronoi cell corners are the triangle circumcenters, which makes cell and triangle edges orthogonal (Fig.~\ref{fig:vertex-models}b). As we will see below, the duality between triangulations and Voronoi tessellations plays an important role in the DDG framework. 
The dynamics of the $\mathbf{r}_i$ comprises two terms, self-propulsion and relaxation of an elastic energy:
\begin{align}
    \label{eq:SVP_dynamics}
    \partial_t \mathbf{r}_i = -\nabla_{\mathbf{r}_i} E_\mathrm{SPV} + v_0 \hat{\mathbf{v}}_i.
\end{align}
For each cell $i$, $\hat{\mathbf{v}}_i$ is a unit vector (represented by an angle $\theta_i$) that determines the direction of motion. 
The elastic energy depends on each cell's Voronoi area $a_i$ and perimeter $p_i$:
\begin{align}
    \label{eq:SVP_energy}
    E_\mathrm{SPV} = \sum_i k_a(a_i-a_0)^2 + k_p(p_i-p_0)^2,
\end{align}
where $k_a, k_p$ are elastic constants, and $a_0, p_0$ are the target area and perimeter. 
The orientation $\theta_i$ of each cell undergoes rotational diffusion,
$d\theta_i = D dW_{t, i} $. Here $dW_{t,i}$ is Brownian motion, independent for each cell $i$, and $D$ is a diffusion constant.
In the SPV, cells can rearrange if the interface between two cells shrinks to length 0, leading to an edge flip in the triangulation (Fig.~\ref{fig:vertex-models}d).

To simulate Eq.~\eqref{eq:SVP_dynamics}, we implemented the elastic energy Eq.~\eqref{eq:SVP_energy} using the \texttt{geometry} module, and used autodiff to compute its gradient. To integrate the stochastic differential equation Eq.~\eqref{eq:SVP_dynamics}, we used the \texttt{diffrax} library. After each time step, we used the \texttt{topology} module to carry out edge flips. 
The entire simulation can thus be accelerated using  \jx's JIT-compilation. 
A snapshot of 8 cells from one such simulation is provided in Fig.~\ref{fig:vertex-models}c.

\subsubsection{Differentiating simulations with stochastic forces and topological changes}

Since all \tx operations are differentiable, one can differentiate the output of entire simulations with respect to model parameters. 
This allows optimizing parameters to fit experimental data or to achieve a design objective (make the model carry out a desired behavior). Autodiff works even in simulations that involve stochastic forces and topological modifications (e.g., T1 transitions in vertex models). Intuitively, T1 transitions seem non-differentiable: they are discrete changes of the mesh connectivity. However, the Voronoi model evolves continuously through a T1 transition, because the cell edge has zero length at the moment of the flip.
Stochastic forces pose no problem for autodiff, either: the noise does not depend on simulation parameters and drops out from the gradient.

\jx computes the gradient in which the stochastic forces and the sequence of T1 transitions are kept fixed. 
Although this gradient is technically not the exact derivative of the simulation output, it suffices for many practical purposes.
For example, the autodiff gradient agrees with finite differences and is effective for optimizing model parameters. 
This is demonstrated in Fig.~\ref{fig:topo autodiff supp} with a toy example, in which we optimize the SPV parameters to achieve target tissue and cell shapes. 
Autodiff reliably computes the gradient for simulations with $\geq 1,000$ T1 transitions. However, the accuracy of gradients decreases for long simulation runs if the cell dynamics are chaotic, a well-known issue in (reinforcement) machine learning~\cite{sutton1998reinforcement}.

\subsection{Partial differential equations on surfaces}

\begin{figure*}
    \centering
    \includegraphics{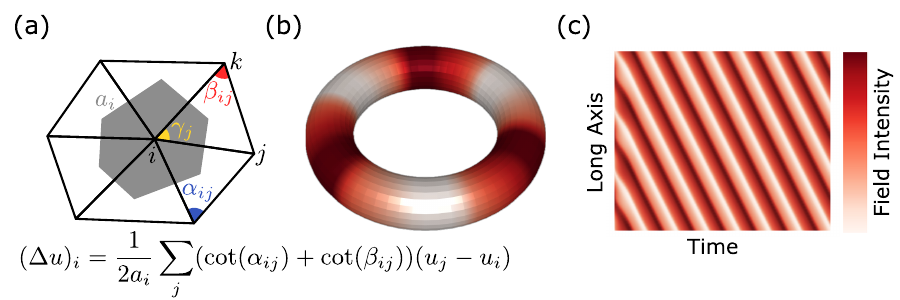}
    \caption{\textbf{Partial differential equations on curved surfaces.} 
    (a) The discrete Laplace operator on triangular meshes is defined by the cotan Laplacian.
    (b) NRCH model on a torus embedded in 3d leads to traveling waves along the long axis. Parameter values: $D=-0.14, d= 0.1, \int \phi dA = \int \psi dA = 0$.  
    (c) 2d kymograph along the long axis on the outer radius of the torus in (b).
    }
    \label{fig:patterns}
\end{figure*}

To simulate PDEs on curved and potentially dynamic surfaces, one can use \tx's discrete differential operators, like the Laplace operator.
The DDG discretization combines aspects of linear finite-element and finite-volume methods, with modifications to improve stability on irregular meshes~\cite{Meyer.etal2003}.
Continuous fields $u(\mathbf{x})$ are approximated by their values $u_i$ at vertices, and the discrete Laplace operator is a sparse matrix, referred to as the cotan-Laplacian:
\begin{align}
    \label{eq:cot_formula}
    (\Delta u)_i = \frac{1}{2 a_i} \sum_{j} (\cot\alpha_{ij} +\cot\beta_{ij}) (u_j-u_i),
\end{align}
where the sum is over all neighbors $j$ of vertex $i$, $a_i$ is the Voronoi area around vertex $i$, and the angles $\alpha, \beta$ are defined in Fig.~\ref{fig:patterns}. 
Eq.~\eqref{eq:cot_formula} follows from the divergence theorem, applied to the Voronoi cell around each vertex, and is equivalent to the Laplacian obtained from the linear Lagrange elements in the finite-element method~\cite{Crane.etal2013}.
For static surfaces, \tx thus produces identical results to finite-element simulations. We calculate $a_i$ robustly using the method of Ref.~\cite{Meyer.etal2003}.

Using the cotan Laplacian, we simulated a nonreciprocal version of the Cahn--Hilliard model (NRCH)~\cite{Brauns.Marchetti2024}, a generic model of pattern formation and traveling waves, on a torus embedded in 3d (Fig.~\ref{fig:patterns}). The NRCH comprises two conserved fields $\phi, \psi$, evolving as:
\begin{align}
    \label{eq:NRCH}
    \partial_t \phi = \Delta \left(-\phi + D\psi + \phi^3 - \Delta\phi\right), \quad \partial_t \psi = \Delta \left(|D|\phi + d\,\psi\right),
\end{align}
where $D, d$ are diffusion coefficients. We discretized Eq.~\eqref{eq:NRCH} using the cotan Laplacian and an implicit Euler scheme. For analysis, \tx also supports projecting data from 3d to 2d using coordinate charts (known as ``UV maps'' in computer graphics).

\subsection{Membrane elasticity}\label{sec:membrane_elasticity}

Membranes are often modeled as 2d surfaces embedded in 3d~\cite{Seifert1997}.
Here, we consider a membrane for which the surface tension is negligible and the elastic energy is dominated by bending. This is the case for many lipid bilayers, like the cellular plasma membrane.
Such lipid bilayers combine aspects of fluids and solids: they resist out-of-plane bending elastically, but flow viscously in-plane~\cite{Seifert1997}.
The \emph{Helfrich energy} is a geometric model for the bending energy of a surface~\cite{Helfrich1973}, based on the Gaussian and mean curvatures $K, H$:
\begin{align}
    \label{eq:Helfrich}
    E_\mathrm{H}  =\int dA \left( \frac{k}{2}(H-H_0)^2 + k' K \right),
\end{align}
where $k, k'$ are elastic moduli and $H_0$ is the preferred (mean) curvature of the membrane. For closed surfaces, the second term is a topological constant and is thus irrelevant for simulations of meshes with fixed topology. Much of the physics of membranes arises from constraints on the volume $V$ and area $A$ of the membrane, which force the adoption of non-trivial bent shapes. We describe these constraints via the reduced volume $v = 6\sqrt{\pi} \, V / A^{3/2}$ relative to a sphere with the same area~\cite{Seifert.etal1991}.

\subsubsection{Discretized elastic energy}

\tx discretizes this model using DDG. The Gaussian curvature at vertex $i$ becomes the \emph{angular defect} $K_i = (2\pi - \sum_{j} \gamma_{j})/a_i$, where the sum is over all angles $\gamma_j$ at vertex $i$ (Fig.~\ref{fig:patterns}a).
DDG also provides a discrete expression for the mean curvature $H_i$ at each vertex (Eq.~\eqref{eq:steiner} in SI.~\ref{sec:DDG_SI}).
Thus, all terms in Eq.~\eqref{eq:Helfrich} are computed without the higher-order derivatives required in coordinate parametrizations.
The area integral $\int dA \left(\cdot \right)$ becomes $\sum_i a_i \left(\cdot \right)$, so that the discretized energy reads
\begin{align}
    E_H = \sum_i a_i \left(\frac{k}{2}(H_i-H_0)^2  + k' K_i\right) .
\end{align}
The resulting force density $\mathbf{f}_i$ per vertex is
\begin{align}
    \mathbf{f}_i = -\frac{1}{a_i}\frac{\delta E}{\delta \mathbf{r}_i} .
\end{align}
The functional derivative $\tfrac{\delta E}{\delta \mathbf{r}_i}$ (computed via autodiff) represents the total force, integrated over the area around a vertex (vertices with larger area influence $E$ more strongly). In finite-element terms, the Voronoi areas $a_i$ define the mass matrix of the discretization.

\subsubsection{Numerical optimization}
\label{sec:Numerical optimziation}

Using the discretized bending energy, we obtained minimum-energy membrane shapes by gradient-based optimization, using \jx to automatically compute derivatives. We used the \texttt{optimistix} library of non-linear optimizers~\cite{optimistix2024}.
We implemented area and volume constraints using the augmented Lagrangian method~\cite{Nocedal.Wright1999}.

\begin{figure*}
    \centering
    \includegraphics[width=0.8\textwidth]{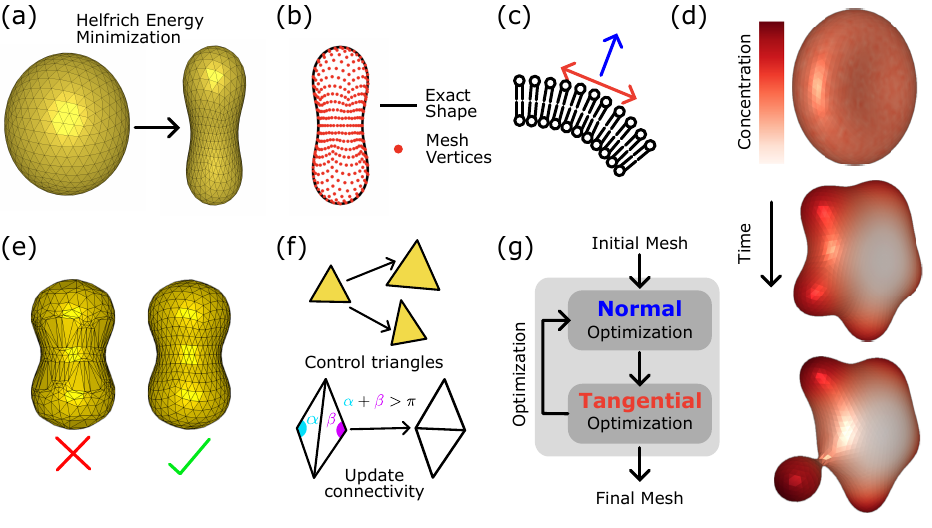}
    \caption{\textbf{Membrane mechanics.}
    (a) Initial condition (left) and minimum-energy shape for the Helfrich energy Eq.~\eqref{eq:Helfrich} with reduced volume $v=0.75$.
    (b) The results from \tx match the exact, rotationally symmetric solution.
    (c) Membranes are lipid bilayers that elastically resist out-of-plane bending (red), but behave like a viscous fluid in-plane (blue).
    (d) Coupling to concentration fields can lead to instability and geometric pattern formation. Snapshots from relaxational dynamics of Eq.~\eqref{eq:leibler} with $k =1$, $\Lambda=2$, $a=1$, $b = 1$. The initial condition is an (oblate) minimum energy shape for $v=0.75$.
    (e) Membrane energy only depends on surface shape, not the in-plane position of vertices.
    Many meshes may represent the same shape, but some meshes are more preferable to others. Skinny triangles can lead to numerical instability in simulations, so uniform equilateral triangulations are preferable.
    (f) Mesh quality is improved by controlling triangle shapes (size, aspect ratio) by optimizing an in-plane pseudo-energy, and by performing edge flips for poorly conditioned triangles. Both operations preserve the number of vertices and are fully compatible with \jx.
    (g) Motion of mesh vertices is split between a normal phase for the physical energy and a tangential phase for the parameterization energy.
    }
    \label{fig:membranes}
\end{figure*}

Numerical optimization of Eq.~\eqref{eq:Helfrich} is notoriously challenging due to its \emph{reparametrization} invariance~\cite{Brakke1992}.
Membranes are fluid and do not elastically resist in-plane material redistribution, only out-of-plane bending. In the discrete setting, moving vertices in the local tangent plane does not change the energy, which only depends on the surface shape. 
This means the optimization landscape has flat directions that cause mesh degeneration, leading to numerical instability.
Inspired by Ref.~\cite{Sahu2025}, we address this issue by splitting vertex motion into two sub-problems:
\begin{enumerate}
    \itemsep-.3em 
    \item \emph{Normal phase}: minimize the physical energy Eq.~\eqref{eq:Helfrich} by motion along surface normals.
    \item \emph{Tangential phase}: minimize a parametrization pseudo-energy by motion orthogonal to surface normals.
\end{enumerate}
The parametrization pseudo-energy penalizes degenerate mesh configurations, and the restricted optimization only along the tangential direction ensures that it does not bias the physical minimum-energy shapes.
We use a \emph{neo-Hookean} parametrization energy (see next section), but this choice can easily be switched out, since gradients (forces and stresses) are computed automatically.
Iteratively minimizing normal and tangential energy, we obtain minimum energy shapes with low mesh distortion (Fig.~\ref{fig:membranes}a). We verified our results against analytical solutions for rotationally symmetric surfaces, computed by the method of Ref.~\cite{Seifert1997} (Fig.~\ref{fig:membranes}b; see SI Sect.~\ref{sec:rotational symmetry check}).

\subsubsection{Coupling to scalar patterns}

Many substances, including large classes of proteins~\cite{Stachowiak.etal2012}, can bind to membranes and modify their preferred curvature $H_0$ \cite{Seifert1997}.
Shape-concentration coupling is of great importance for sculpting the shapes of cells and intracellular organelles~\cite{Gov2018}.
We simulate a minimal model for a shape instability induced by this geometry-concentration coupling~\cite{Leibler1986}.
We consider a single scalar concentration field $\phi$ restricted to the membrane surface, whose Landau free energy of the concentration field around the equilibrium value $\phi=0$ is $\int dA \left( \frac{a}{2} \phi^2 + \frac{b}{2} |\nabla \phi|^2 \right)$.   
We assume $\phi$ induces a spontaneous curvature $H_0 = \Lambda \phi$.
Thus, the joint free energy of membrane and field reads
\begin{align}
    \label{eq:leibler}
    F = \int dA \left(\frac{k}{2}H^2 - \Lambda H\phi + \frac{a}{2} \phi^2 + \frac{b}{2} |\nabla \phi|^2  \right) .
\end{align}
Eq.~\eqref{eq:leibler} implies that, vice versa, the $\phi$-field ``prefers'' regions of non-zero curvature. 
For sufficiently large $\Lambda$, this leads to an instability (high-$\phi$ regions increase in curvature, attracting yet more $\phi$).

We used \tx to simulate the overdamped dynamics of this model\footnote{Strictly speaking, these pseudo-dynamics minimize the free energy but do not capture $\phi$-advection by in-plane flows.}. Both membrane shape (vertex positions $\mathbf{r}_i$) and the field (values $\phi_i$) relax the free energy,
\begin{align}
    \partial_t \phi = -\delta F / \delta \phi, \quad \partial_t \mathbf{r} = -\delta F / \delta \mathbf{r}.
\end{align}
These relaxational dynamics, in which $\phi$ is not conserved, are chosen as a minimal example (see Discussion). Conserved dynamics ($\partial_t\phi=\Delta \delta F/\delta \phi$) are also straightforward to implement using the cotan-Laplacian.
\tx automatically computes these functional derivatives from the discretized version of Eq.~\eqref{eq:leibler}. Fig.~\ref{fig:membranes}d shows the development of the resulting instability. Here, we enforced the volume and area constraints using a quadratic energy penalty.
 
\subsection{Elasticity of thin shells}

Elastic deformations of thin shells are an important subject in soft-matter physics, which can be elegantly described using the framework of metric elasticity~\cite{Efrati.etal2013,VanRees.etal2017}.
In brief, the physical shape of an elastic shell with thickness $h$ is described by a Riemannian metric $g$, measuring the physical distance between nearby material points. The reference configuration corresponds to a second metric $g_r$, measuring the energetically preferred distance between material points (Fig.\ref{fig:elastic}a).
The elastic energy is computed from the non-linear deformation tensor $C= g_r^{-1}\cdot g$. A common choice is the neo-Hookean model:
\begin{align}
    \label{eq:neo-hookean}
\begin{aligned}
    E_{\mathrm{nH}} = h\int dA \Bigg(&\frac{G_s}{2}\left(\frac{\mathrm{tr}(C)}{\sqrt{\det C}} - 2\right)
 \\
  &+\frac{G_b}{2}\left(\sqrt{\det C} - 1\right)^2 \Bigg),
\end{aligned}
\end{align}
where $G_s, G_b$ are the shear and bulk moduli. 
Due to its favorable stability behavior even for large deformations, we used Eq.~\eqref{eq:neo-hookean} as the parametrization pseudo-energy in the membrane simulations above.
To implement Eq.~\eqref{eq:neo-hookean} discretely, we followed Refs.~\cite{Grinspun.etal2003,Chen.etal2018} and represented the metric tensor $g_{ijk}$ of each triangle $ijk$ intrinsically in terms of the triangle edge vectors (SI Sect.~\ref{sec:DDG_SI}).
The energy Eq.~\eqref{eq:neo-hookean} only describes the energy due to in-plane deformations. In addition to the metric $g$, capturing bending requires the $2^\mathrm{nd}$ fundamental form $b$, which measures the extrinsic curvature through the change of the surface normal $\mathbf{n}$ (the mean curvature is $2H = \mathrm{tr}\, g^{-1} b$). A simple bending energy reads~\cite{Grinspun.etal2003}
\begin{align}
    \label{eq:bending}
    E_{\mathrm{B}} = G_f\int dA \,\mathrm{tr}((b-b_r)^2) .
\end{align}
$G_f \propto h^3$ is the flexural modulus, and $b_r$ is the reference $2^\mathrm{nd}$ fundamental form. A simple way to generate a curved reference configuration is to glue together two sheets with different target metrics into a bilayer (e.g., an expanding top and contracting bottom sheet)~\cite{VanRees.etal2017}. 
\tx also implements a discrete version of Eq.~\eqref{eq:bending}, based on Ref.~\cite{Grinspun.etal2003}. To discretize $b$, one computes the dihedral angles $\varphi_{ij}$ between the normal vectors of the two triangles that meet at an edge $ij$ (Fig.\ref{fig:elastic}a, top). The bending energy becomes  
\begin{align}
    \label{eq:bending_discrete}
    E_{\mathrm{B}} = B \sum_{ij} \frac{e_{ij}^2}{a_{ij}} (\varphi_{ij} -\varphi_{ij, r})^2 ,
\end{align}
where $e_{ij}$ is the edge length, and $a_{ij}$ is the edge area ($1/3$ of the summed area of the two adjacent triangles). \tx also implements a generalization of Eq.~\eqref{eq:bending_discrete} from Ref.~\cite{Chen.etal2018} in which $b_r$ is encoded as a per-face tensor instead of target angles $\varphi_{ij, r}$. 

\subsection{Inverse design of a shape-changing elastomer sheet}

\begin{figure*}[t]
    \centering
    \includegraphics[width=1\textwidth]{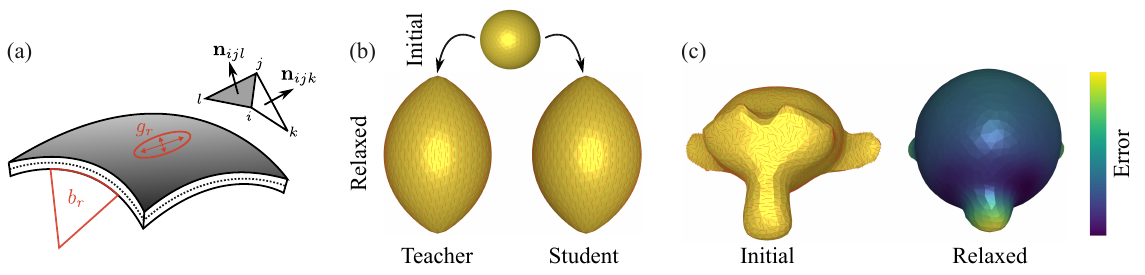}
    \caption{\textbf{Thin shell elasticity and inverse problems.} 
    (a) In metric elasticity, a thin elastic sheet has a reference metric $g_r$, specifying the minimum-energy distance between nearby material points, and a reference $2^\mathrm{nd}$ fundamental form, specifying the minimum-energy bending radii and axes. 
    On a discrete surface, bending is measured by the dihedral angle between adjacent triangle normals.    
    (b) An initial spherical mesh relaxes to an elongated shape when the ``teacher'' nematic pattern is actuated, deforming the reference metric. Red line segments show the nematic director $\theta_t$. After training, the student recovers the teacher's shape and pattern. Parameter values: $G_s=G_b=G_f=1, h=0.1, \lambda=1.1$.
    (c) Nematic pattern trained to deform a monkey's head into a sphere upon actuation. Deviation from sphericity is measured by the radius variance, which drops by a factor of 10, although the snout and ears are not completely flattened. Parameter values: $G_s=G_b=G_f=1, h=0.1, \lambda=1.5$.
    }
    \label{fig:elastic}
\end{figure*}

Finally, we applied \tx to an example inverse problem, the design of a shape-changing liquid crystal elastomer sheet, following Refs.~\cite{Aharoni.etal2014,Aharoni.etal2018}.
Such sheets have a nematic orientation $\theta$ imprinted. When heated, the material locally shrinks along the nematic director and expands in the orthogonal direction.
The sheet is modeled using metric elasticity, Eq.~\eqref{eq:neo-hookean}. The reference metric of the actuated state is determined by the nematic director:
\begin{align}
    \label{eq:metric_actuated}
    g_r = \Lambda^T \cdot g_0 \cdot \Lambda, \quad \Lambda = R(\theta)\cdot \mathrm{diag}(\lambda, 1/\lambda)\cdot R(\theta)^T,
\end{align}
where $g_0$ is the metric of the initial, non-heated state, $\lambda > 1$ is the material's expansion factor, and $R$ is a rotation matrix. To control the sheet's curvature, one can glue together two sheets with different directors $\theta_t,\theta_b$. Differences in strain across the two layers induce nonzero reference curvature $b_r \neq 0$.

The goal is to paint two nematic director fields onto an initial surface $\mathcal{S}_0$ so that the elastically relaxed shape after heating matches a desired target $\mathcal{S}_1$. Because the expansion factor $\lambda$ is a fixed material constant, the problem is constrained: it is not possible to morph each piece of the material into the target shape.
Note that Ref.~\cite{Aharoni.etal2018} already introduces a numerical algorithm for this inverse problem, based on elegant geometry and computer-graphics techniques.
We propose an autodiff-based approach that, while less efficient, is more easily generalized. It is a black-box approach that works with any forward simulation model and does not exploit special structure of the model.

Given an orientation field $\theta_t,\theta_b$ (encoded as per-triangle angles), we start from $\mathcal{S}_0$ and compute the minimum energy shape of the actuated reference metric Eq.~\eqref{eq:metric_actuated}, resulting in a shape $\hat{\mathcal{S}}_1$. We then measure the difference to the target shape $\mathcal{S}_1 $ using a loss $\mathcal{L}$.
The loss can be augmented by additional penalty terms $\mathcal{P}$, for example, to encourage smooth director fields.
We then minimize $\mathcal{L}$ with respect to the design fields $\theta_t,\theta_b$ to obtain the optimized pattern: 
\begin{align}
    \label{eq:two-level}
    \begin{aligned}
    \theta_t^*,\theta_b^* &= \arg \min_{\theta_t,\theta_b}\mathcal{L}[\hat{\mathbf{r}}_1]+ \mathcal{P}[\theta_t,\theta_b] , \\ \text{with} \; \hat{\mathbf{r}}_1 &= \arg \min_\mathbf{r} E(\mathbf{r}; \theta_t,\theta_b).
    \end{aligned}
\end{align}
Since the \tx simulation is differentiable, minimization can be carried out with gradient-based methods. In traditional, non-differentiable simulations, optimization or fitting requires parameter scans, which become infeasible for high-dimensional problems (here, the design fields have dimension $2\times\text{No. of faces}$).
We use the conjugate-gradient solver from \texttt{optimistix} to minimize the energy, and the Adam optimizer~\cite{Kingma.Ba2014} from the machine-learning library \texttt{optax} to minimize the loss~\cite{deepmind2020jax}. Simulation details (loss functions, optimizer settings) can be found \href{https://nikolas-claussen.github.io/triangulax/}{online}.

As a first example, we paint a ``teacher'' pattern on a spherical mesh that deforms it into an ellipsoid (Fig.\ref{fig:elastic}b). With this deformed shape as target $\mathcal{S}_1$, we use the above method to recover the teacher pattern.
The optimized ``student'' pattern reproduces the teacher shape almost exactly, and recovers the teacher pattern.
As a second example, we aim to morph a complex shape into a sphere (Fig.\ref{fig:elastic}c). We chose a smoothed version of the ``Suzanne'' mesh of a monkey's head from the 3d software Blender. The loss now penalizes the deviation of the relaxed configuration $\hat{\mathcal{S}}_1$ from a sphere.
The optimized pattern comes close to accomplishing this task, dropping the deviation from sphericity by $10\times$ (although the snout and ears are not completely flattened).

\section*{Discussion}

We have presented \tx, a geometry-processing and simulation toolkit for curved 2d surfaces that combines discrete differential geometry~\cite{Crane.etal2013} and automatic differentiation~\cite{jax2018github}.
Based on discrete differential geometry, we developed an algorithm for simulations of surfaces undergoing large deformations without requiring coordinate charts.
The library is based on the \jx framework~\cite{jax2018github} and thereby provides access to key machine-learning technology. Autodiff greatly facilitates both minimization of (multi-physics) free energies and allows optimizing the parameters of entire simulations.
\tx is not ``hard-coded'' for soft-matter applications and can be applied to other physical or engineering systems described by curved surfaces, embedded into spaces of arbitrary dimension and periodicity.

We demonstrated \tx on a range of problems across soft-matter physics, from vertex models~\cite{Farhadifar.etal2007} to membrane mechanics~\cite{Seifert1997}.
We validated our discretizations against exact solutions and existing software.
An important limitation of the example simulations shown in Fig.~\ref{fig:membranes} is that they do not fully model the viscous hydrodynamics of in-plane membrane flows~\cite{Arroyo.DeSimone2009}. We aim to release a complete tutorial with the next iteration of the software, using a stream-function approach that can be implemented using the \tx toolset.
More generally, our simulations implicitly assumed an ``adiabatic limit'' in which the membrane relaxes to mechanical equilibrium rapidly compared to the evolution of the scalar concentration field due to diffusion.

More generally, due to its use of autodiff, \tx is well-suited to ``multiphysics'' simulations, like protein patterns on deforming surfaces (Fig.~\ref{fig:membranes}d). Within the framework of Onsager’s variational principle~\cite{Doi2011,Arroyo.etal2018}, a multiphysics model is specified by the joint free energy and dissipation potential of all variables (concentrations, surface shape, etc.) The dynamics are governed by minimizing the total energy dissipation rate. Autodiff automates the calculation of the resulting (generalized) forces and frictions, and facilitates the dissipation rate minimization. 
\tx-based ``surface'' models can also be integrated with other physics, like particle-based models or 3d bulk fluid dynamics, for example via the immersed boundary method \cite{Peskin1972}. 

In future work, we plan to use \tx to study how concentration patterns, governed by reaction-diffusion dynamics, can sculpt the geometry of fluid membranes, a key question for the shape of cells and organelles~\cite{Gov2018}.
Compared to coordinate-based approaches~\cite{Yu.Kosmrlj2025,Winter.etal2025}, \tx can handle dramatic shape deformations, for example during tubulation or budding.
We further anticipate that the ability to automatically differentiate entire simulations will unlock new possibilities, for example, to develop quantitative models of biological processes that involve complex nonequilibrium dynamics on deformable membranes~\cite{FreemanRosenzweig.etal2017,Hsieh.etal2021,Hsieh2026_apatterns,Yu2026_membrane-anchoredmigration}.

More generally, we believe that gradient-based optimization is a promising route to explore the principles of surface shape programming.
Indeed, in toy models (cellular automata), optimization of differentiable models has generated striking examples of self-organized pattern generation~\cite{Mordvintsev.etal2020}, and shed light on the underlying mechanisms.
We hope that \tx brings this exciting line of research to the wide range of biophysics and soft-matter models that can be simulated with triangular meshes.

\begin{acknowledgments}

We thank Fridtjof Brauns, Andrej Košmrlj, Noah Mitchell, and Ned Wingreen for helpful discussions and feedback on the manuscript.
Research reported in this publication was supported by the National Science Foundation through the grants MCB-2410354 and DMR-2011750 (through the Princeton Institute for Complex Materials), and by the National Institute of General Medical Sciences (NIGMS) of the National Institutes of Health (NIH) under award numbers R01GM082938 and R35GM156427. 
The content is solely the responsibility of the authors and does not necessarily represent the official views of the National Institutes of Health or the National Science Foundation.
S.S. is supported by a Princeton University First-Year Fellowship and a summer research assistantship from the Princeton Physics Department.
The work by Q.Y. is supported in part by an LITP fellowship from the Leinweber Institute for Theoretical Physics at Stanford University.
N.C. is supported by a fellowship from the Princeton Center for Theoretical Science.

\paragraph*{Author contributions}
N.C. designed research and developed software. S.S. and Q.Y. performed research. S.S., Q.Y., and N.C. wrote the manuscript.

\paragraph*{AI usage disclosure}
We used Claude Code by Anthropic (Models Opus 5 and 5.5) for programming assistance and automated software testing.

\paragraph*{Data availability}
The software package, documentation, and tutorials are available online at \href{https://github.com/nikolas-claussen/triangulax}{https://github.com/nikolas-claussen/triangulax}.

\paragraph*{Conflicts of interest} 
The authors declare no conflicts of interest.

\end{acknowledgments}

%\printbibliography

\bibliography{triangulax.bib}

\setcounter{section}{0}
\setcounter{subsection}{0}

\renewcommand{\thesection}{S\Alph{section}}
\renewcommand{\thesubsection}{S\arabic{subsection}}
\makeatletter\renewcommand{\p@subsection}{}\renewcommand{\p@subsubsection}{}\makeatother

\renewcommand{\thesubsubsection}{\thesubsection.\arabic{subsubsection}}

\setcounter{figure}{0}
\renewcommand{\thefigure}{S\arabic{figure}}

\setcounter{equation}{0}
\renewcommand{\theequation}{S\arabic{equation}}

\setcounter{table}{0}
\renewcommand{\thetable}{S.\Roman{table}}

\clearpage
\section*{Supplementary Information}

\begin{table*}[]
    \centering
    \begin{tabular}{l|l}
    \toprule
    Module name & Functionality \\
    \midrule
     \texttt{trigonometry} & trigonometry and linear algebra utilities \\
     \texttt{triangular} & mesh input/output \\
     \texttt{mesh} & half-edge data structure for triangular meshes, compatible with \jx \\
     \texttt{topology} & topological mesh modifications \\
     \texttt{adjacency} & vertex-vertex, vertex-face, and face-face adjacency operators \\
     \texttt{geometry} & angles, edge lengths, triangle areas, Voronoi areas, curvatures \\
     \texttt{periodic} & geometry with periodic boundary conditions \\
     \texttt{linops} & differential operators (Laplacian, gradient, $\dots$) \\
     \texttt{interp} & linear interpolation \\
     \texttt{elastic} & elastic energies for shells and membranes \\
     \texttt{algorithms} & diverse mesh utilities (statistics, quality improvement) \\
     \texttt{simulation} & simulation management \\
    \bottomrule
    \end{tabular}
    \caption{List of modules available in \tx (see \href{https://nikolas-claussen.github.io/triangulax/}{online documentation} for details).}
    \label{tab:modules}
\end{table*}

\subsection{Geometric objects and operators in DDG}
\label{sec:DDG_SI}

Our summary here follows Ref.~\cite{Crane.etal2013}. The discretization described below matches the implementation in \tx.

\paragraph{Discrete exterior calculus (DEC)}
Exterior forms are fundamental to differential geometry. On a 2d manifold, there are three types: 0-forms (ordinary functions), 1-forms (covectors), and 2-forms (scalar densities). A $k$-form can be integrated over a $d-k$-dimensional subspace. Thus, DEC discretizes continuous $k$-forms by integrating them over $d-k$-dimensional mesh elements~\cite{Desbrun.etal2005}. A function $u$ is represented by its vertex values $u_i =u(\mathbf{r}_i)$, a covector field $\boldsymbol{\alpha}$ by the integral over edges $\alpha_{ij} = \int_{ij} \boldsymbol{\alpha} \cdot d\mathbf{e}$, and a density $\rho$ by the integral over faces, $\rho_{ijk} = \int_{ijk} \rho \, dA$. Note that both faces and edges are oriented.

The exterior derivative $d$ maps $k$ to $k+1$-forms (for example, on $1$-forms, $d$ corresponds to the familiar curl operator). It obeys Stokes' theorem, $\int_{\partial \Omega}\alpha = \int_{\Omega} d\alpha$, where $\Omega$ is an integration region and $\alpha$ a $k$-form. In DEC, Stokes' theorem defines the discrete version of $d$. For example, the discrete gradient is $(du)_{ij} = \int_{\partial(ij)} u = u(\mathbf{r}_j) - u(\mathbf{r}_i)$ and the discrete curl around a triangle is $(d\boldsymbol{\alpha})_{ijk} = \boldsymbol{\alpha}_{ij} + \boldsymbol{\alpha}_{jk} + \boldsymbol{\alpha}_{ki}$.

\paragraph{Area integral, Laplace operator, and Hodge star}

In differential geometry, the metric $g$ defines the local area element $\sqrt{\det g} \,d^2x$. To assign an area to each vertex $\mathbf{r}_i$, we use Voronoi duality (Fig.~\ref{fig:vertex-models}), which assigns a polygonal cell to each vertex. The cell corners are the triangle circumcenters, which makes cell and triangle edges orthogonal.
The Voronoi area $a_i$ defines the local area element, leading to the discretization $\int(\cdots) \sqrt{\det g} \, d^2 x \mapsto \sum_i (\cdots)_i \, a_i$ of the area integral. We implement a version of $a_i$ that is robust on irregular meshes~\cite{Meyer.etal2003}.
\tx also provides the barycentric vertex areas (the average of the surrounding triangle areas), used by many implementations instead of the Voronoi areas~\cite{Zhu.etal2022}. However, the Voronoi areas lead to more accurate discretizations~\cite{Meyer.etal2003}.

Applying the divergence theorem to the Voronoi cell $i$ leads to a discrete Laplace operator:
\begin{align}
    \begin{aligned}
    (\Delta u)_i &:= \frac{1}{a_i} \int_i \Delta u \,dA = \frac{1}{a_i}\int_{\partial_i} \nabla u\cdot d \mathbf{n} \\
    &\approx 
    \frac{1}{a_i}\sum_{j} e_{ij}^* \frac{u_j-u_i}{e_{ij}}.
    \end{aligned}
\end{align}
The ratio of cell $e_{ij}^*$ and triangle edge lengths $e_{ij}$ works out exactly to the cotangent weights in Eq.~\eqref{eq:cot_formula}.
More generally, DDG uses the Voronoi duality to define a discrete equivalent of the differential-geometric Hodge star operator~\cite{Desbrun.etal2005}. The Hodge star uses the Voronoi cell areas $a_i$ and cell edge lengths $e^*_{ij}$, available in \tx.

\paragraph{Interpolation, divergence, gradient, and link to finite-element method}

Evaluating a discrete function defined by vertex values $u_i$ at arbitrary points on the surface requires interpolation. It is convenient to introduce a basis of finite-element test functions~\cite{Brenner.Scott2008}. These $\phi_i(\mathbf{r})$ defined by their values at vertices $\phi_i(\mathbf{r}_j)=\delta_{ij}$ and linear interpolation in between. In terms of this basis, the interpolated function reads $u(\mathbf{r}) = \sum_i u_i \phi_i(\mathbf{r})$. 

The gradient of the test functions is constant on each face and reads $\nabla\phi_i =  \sum_{cyc}(\mathbf{r}_k-\mathbf{r}_j)^\perp / (2a_{ijk})$, where ($^\perp$) denotes a $90^\circ$ rotation about the triangle normal. Thus, the discrete gradient of an interpolated function is constant on each face and reads $(\nabla u)_{ijk} = \sum_{l=i,j,k} u_l \nabla\phi_l$.
The gradient vector $\nabla u$ is expressed in the basis of the 2d or 3d embedding space.
Indeed, alternatively to DEC, vector fields on a surface embedded in $d$-dimensional space can be considered as a set of $d$ scalar functions, the vector's components in Euclidean embedding space. 
The divergence operator on such vector fields is defined by $(\nabla \cdot \mathbf{v})_i = -\sum_{jk} a_{ijk} \mathbf{v}_{ijk} \cdot \nabla \phi_i$, where the sum is over all faces adjacent to vertex $i$.

Composing the discrete gradient and divergence results in the discrete Laplace operator Eq.~\eqref{eq:cot_formula} (cotan Laplacian), $(\Delta u)_i = (\nabla \cdot \nabla u)_i$. Hence, the cotan Laplacian coincides with a finite-element discretization. The Voronoi area weights are referred to as ``mass matrix'' in finite elements.

\paragraph{Normal vectors and curvature}

Normal vectors are naturally defined for each triangle by $\mathbf{n}_{ijk} = \mathbf{e}_{ij}\times\mathbf{e}_{ik} / |\mathbf{e}_{ij}\times\mathbf{e}_{ik}|$. Vertex normals are the average of the surrounding triangle normals, weighted by triangle areas.

Differential geometry distinguishes intrinsic curvature and extrinsic curvature. In 2d, the former is fully defined by the Gaussian curvature $K$. Curvature modifies the angle sum of a triangle $T$ as $\sum_{i=1,2,3} \gamma_i = \pi + \int_T K dA$. Therefore, DDG discretizes the $K$ via the angle deficit at a vertex, $K_i = (2\pi - \sum_{j} \gamma_j) / a_i$.
Extrinsic curvature (bending) is measured by the change in the unit normal $\mathbf{n}$ across the surface. In the discrete setting, this change is measured by the dihedral angle $\varphi_{ij}$ between the face normals that meet at an edge. The mean curvature is defined by $2H = -\nabla\cdot\mathbf{n}$. This is equivalent to $2H\mathbf{n} = -\Delta \mathbf{r}$, where $\mathbf{r}$ is the surface embedding. In DDG, $H$ can thus be discretized by applying the cotan Laplacian (coordinate-wise) to the vertex positions:
\begin{align}
    \label{eq:laplace_H}
    H_i = -\tfrac12 \mathbf{n}_i \cdot (\Delta \mathbf{r})_i
\end{align}
Eq.~\eqref{eq:laplace_H} implicitly contains a weighted combination of the dihedral angles. An alternative discretization, Steiner's formula, uses the dihedral angles directly:
\begin{align}
    \label{eq:steiner}
    H_i = \frac{1}{4a_i} \sum_{j} e^*_{ij} \varphi_{ij} 
\end{align}
\tx implements both Eq.~\eqref{eq:laplace_H} and~\eqref{eq:steiner}.

\paragraph{Tangent basis and fundamental forms}

In coordinate-based geometry, the coordinate derivatives $\mathbf{e}_1=\partial/\partial x_1, \mathbf{e}_2 = \partial/\partial x_2$ define a basis of the local tangent plane. 
DDG defines a basis of the tangent plane to each triangle face $ijk$ using its edge vectors. One choice uses the first and second edge vector, $\mathbf{e}_{1}=\mathbf{r}_j-\mathbf{r}_i, \mathbf{e}_2=\mathbf{r}_k-\mathbf{r}_i$.  One can also define an orthogonal basis, $\hat{\mathbf{e}}_{1}=(\mathbf{r}_j-\mathbf{r}_i)/|\mathbf{r}_j-\mathbf{r}_i|, \hat{\mathbf{e}}_2=\mathbf{n}_{ijk}\times \hat{\mathbf{e}}_{1}$. 
\tx also defines a basis for the tangent plane of each vertex $i$,  defined to be orthogonal to the vertex normal $\mathbf{n}_i$. The first basis vector $\hat{\mathbf{e}}_{i, 1}$ is aligned with an arbitrary edge originating from $i$, projected onto the tangent plane, and $\hat{\mathbf{e}}_{i, 2} = \mathbf{n}_i \times \hat{\mathbf{e}}_{i, 1}$.

In the non-orthogonal $\mathbf{e}_1, \mathbf{e}_2$ face basis, the local metric tensor reads~\cite{Chen.etal2018}
\begin{align}
    g_{ijk} = \begin{pmatrix} |\mathbf{e}_{1}|^2 & \mathbf{e}_{1}\cdot\mathbf{e}_{2} \\   \mathbf{e}_{1}\cdot\mathbf{e}_{2} & |\mathbf{e}_{2}|^2 \end{pmatrix}.
\end{align}
The $2^\mathrm{nd}$ fundamental form measures the change in the normal across the surface. Its discretization uses the edge normals $\mathbf{n}_{ij} = (\mathbf{n}_{ijk}+\mathbf{n}_{ijl}) / |\mathbf{n}_{ijk}+\mathbf{n}_{ijl}|$, the averages of the normals of the two faces $ijk, \, ijl$ that meet at edge $ij$. The formula reads~\cite{Chen.etal2018}
\begin{widetext}
\begin{align}
    b_{ijk} = \begin{pmatrix}
    2\mathbf{e}_{1}\cdot(\mathbf{n}_{jk}-\mathbf{n}_{ki})
    & \mathbf{e}_{1}\cdot(\mathbf{n}_{jk}-\mathbf{n}_{ij}) + \mathbf{e}_{2}\cdot(\mathbf{n}_{jk}-\mathbf{n}_{ki})
    \\ \mathbf{e}_{1}\cdot(\mathbf{n}_{jk}-\mathbf{n}_{ij}) + \mathbf{e}_{2}\cdot(\mathbf{n}_{jk}-\mathbf{n}_{ki})
    & 2\mathbf{e}_{2}\cdot(\mathbf{n}_{jk}-\mathbf{n}_{ij}) \end{pmatrix}.
\end{align}
\end{widetext}

\paragraph{Mesh volume} The volume $V$ of a closed mesh can be directly calculated from the vertices $\mathbf{r}_i$ by summing the (oriented) volumes of tetrahedra formed from all mesh triangles and an arbitrary base point. The gradient of $V$ is computed using the custom gradient $d V = a_i \mathbf{n}_i \cdot d\mathbf{r}_i$ instead of using autodiff to remove certain numerical artifacts.

\begin{figure}
    \centering
    \includegraphics[width=1\linewidth]{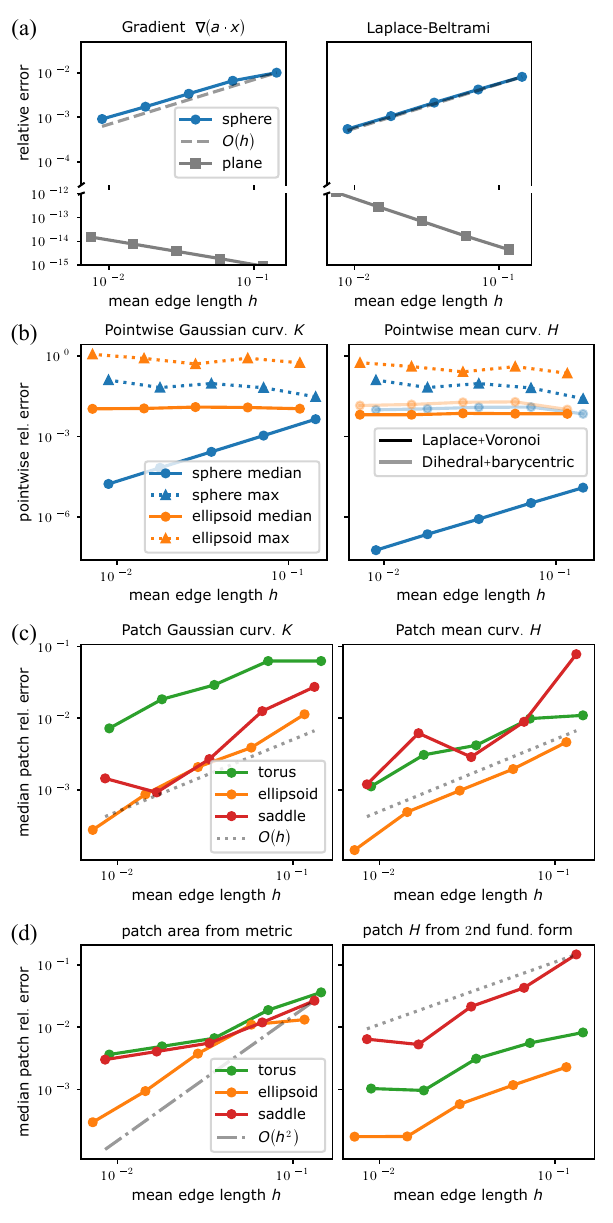}
    \caption{\textbf{Convergence of mean curvature, Gaussian curvature, and Laplace operator with mesh refinement.} 
    Operators were tested on the plane, the unit sphere, an ellipsoid with axis stretch factors $(1.0, 0.8, 0.6)$, a paraboloid $z=\tfrac12(x^2-y^2)$, and a torus with major/minor axis ratio $4$. Meshes of resolution with $h =0.25, 0.5, 1, 2, 4$ percent of the surface bounding box diagonal were generated using the \texttt{MeshLab} software~\cite{Cignoni.etal2008}.
    To test weak convergence (panels c,d), each surface was broken into $8$ disjoint patches.
    Exact values were computed using the formulas for implicit surfaces~\cite{Goldman2005}.}
    \label{fig:convergence}
\end{figure}

\begin{figure*}[tbhp]
\begin{pythonbox}
import jax
import jax.numpy as jnp
from triangulax import triangular, mesh, geometry

# Load example mesh and convert to half-edge mesh

vertices, faces = triangular.read_obj("test_meshes/disk.obj", dim=2)
hemesh = mesh.HeMesh.from_triangles(vertices.shape[0], faces)

# With the half-edge mesh, one can carry out various operations, for example
# computing the coordination number by summing incoming half-edges per vertex:

coord_number = jnp.zeros(hemesh.n_vertices)
coord_number = coord_number.at[hemesh.dest].add(jnp.ones(hemesh.n_hes))
print("Mean coordination number:", coord_number.mean())

>>> Mean coordination number: 5.40458
\end{pythonbox}

\begin{pythonbox}
# Define a simple geometric function and compute its gradient with JAX

def mean_voronoi_area(vertices: jax.Array, hemesh: mesh.HeMesh) -> jax.Array:
    """Compute the mean Voronoi area per vertex."""
    voronoi_areas = geometry.get_voronoi_areas(vertices, hemesh)
    return jnp.mean(voronoi_areas)

value, gradient = jax.value_and_grad(mean_voronoi_area)(vertices, hemesh)
print("Mean gradient norm:", jnp.linalg.norm(gradient, axis=1).mean())

>>> Mean gradient norm: 0.0003638338
\end{pythonbox}
    \caption{\tx usage example.}
    \label{sec:code_example}
\end{figure*}

\subsection{Discretization error and convergence}
\label{sec:convergence}

Like with any numerical scheme, it is important to know if and how the DDG discretizations converge to their smooth counterparts as the mesh resolution (mean edge length $h$) is increased.
Mathematical convergence guarantees exist for the DDG versions of the Laplace operator, mean curvature, and metric~\cite{Hildebrandt.etal2006}, as well as the Gaussian curvature~\cite{Cohen-Steiner.Morvan2003}. Convergence under refinement is only guaranteed if the mesh quality (e.g., triangle deformation) is preserved. If triangles become degenerate, even the area of a surface may not converge to its continuum counterpart~\cite{Marx2016}.

Importantly, like in the finite-element method, one must be careful to distinguish two possible ``modes'' of convergence. Pointwise convergence means that the discrete approximation $Q_h$ of a quantity $Q$ converges at every surface point $p$, $Q(p)-Q_h(p)\rightarrow 0$. Many DDG discretizations, however, only converge in the weak sense, that is, averaged over (arbitrarily small) ``test regions'', $\int_\Omega (Q_h-Q) dA \rightarrow 0$.
Note that the energies of membranes and shells are precisely such surface integrals over energy densities.
Importantly, however, the pointwise DDG discretizations $Q_h(p)$ are stable and do not blow up as the mesh is refined~\cite{Meyer.etal2003}.

In Fig.~\ref{fig:convergence}, we test the convergence of the DDG discretizations for a series of surfaces defined implicitly by an equation $F(x, y, z)=0$, for which the geometric quantities can be calculated exactly~\cite{Goldman2005}.
The gradient and Laplace operators converge linearly pointwise.
To assess convergence in the weak sense, we integrate the quantities over a set of patches whose size remains fixed as the mesh is refined.
As expected, we find that mean and Gaussian curvatures converge linearly, while the metric (area element) converges quadratically.
Gaussian and mean curvature do not converge pointwise, but remain stable. An alternative discretization of the mean curvature using barycentric areas~\cite{Zhu.etal2022} is less accurate than the Laplacian-based formula.

\begin{figure}[tbh]
    \centering
    \includegraphics[width=\linewidth]{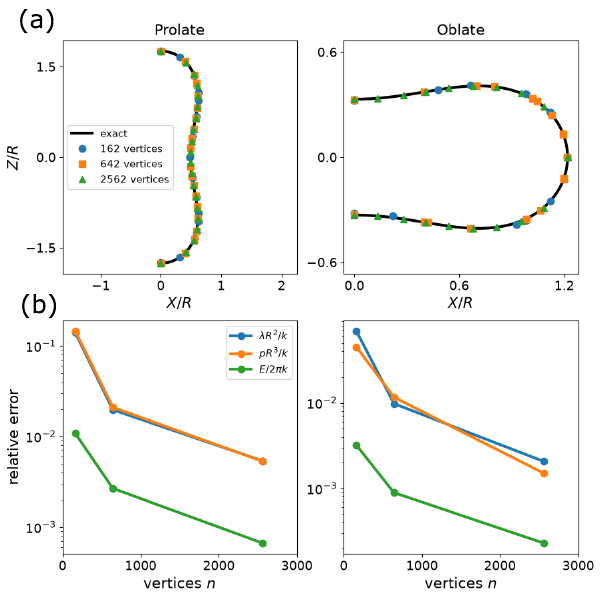}
    \caption{\textbf{Comparison of exact minimal Helfrich energy shapes with results of triangular mesh optimization.} (a) Exact rotationally symmetric profiles for minimal Helfrich energy shapes, compared with vertices of triangular meshes after nonlinear optimization of discretized Helfrich energy. Profiles are reported for 3 levels of mesh refinement with (only 20 vertices shown for clarity). Shapes correspond to $v=0.75$, $R=1$, $H_0=0$ for both prolate (left) and oblate (right) branches. (b) Convergence of (nondimensional) Helfrich energy and Lagrange multipliers (membrane tension $\lambda$ and osmotic pressure difference $p$) to their exact values with mesh refinement.}
    \label{fig:rotational symmetry}
\end{figure}

\begin{figure}[tbh]
    \centering
    \includegraphics[width=\linewidth]{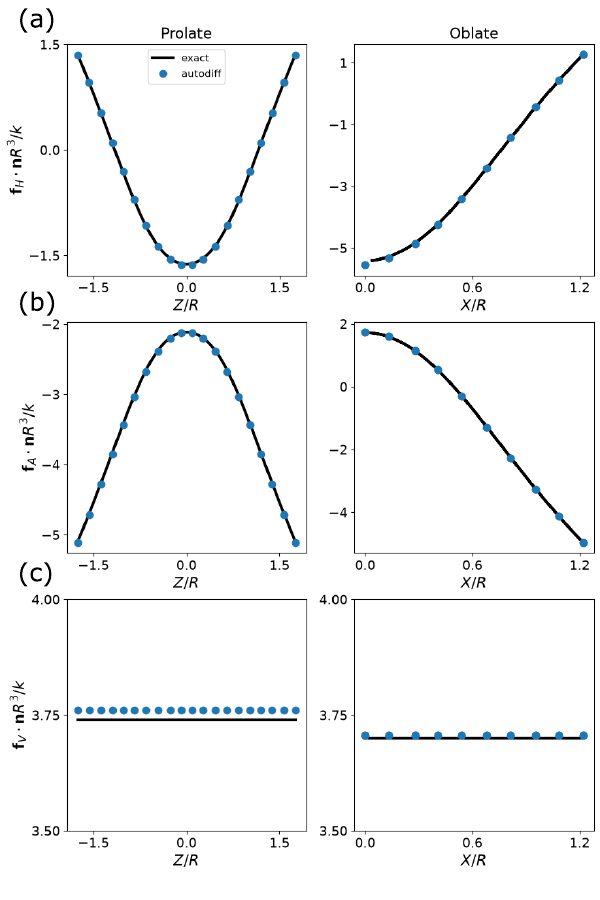}
    \caption{\textbf{Autodiff-derived force densities compared to exact values.} Force densities found by applying autodiff to mesh-based energies compared to exact values for (a) Helfrich energy, (b) constant membrane tension, and (c) constant osmotic pressure. The shapes correspond to the same prolate and oblate shapes as in Fig.~\ref{fig:rotational symmetry}, with autodiff results reported for the 5120 triangle mesh. Only 20 vertices are shown from the triangulated mesh for visual clarity. Forces are densities reported as nondimensional values $\mathbf{f}R^3/k$.}
    \label{fig:autodiff forces}
\end{figure}

\subsection{Minimal Helfrich Energy Shapes With Rotational Symmetry}
\label{sec:rotational symmetry check}

We use rotationally symmetric shapes to verify the minimal Helfrich energy shapes found with numerical optimization of triangular meshes.
Following the method outlined in \cite{Seifert.etal1991}, in which shapes are described by distances along ($Z$) and perpendicular ($X$) to an axis of rotational symmetry, we calculated the profile, energy, and Lagrange multipliers (membrane tension $\lambda$ and osmotic pressure $p$) for minimal Helfrich energy shapes subject to volume and area constraints.

Fig.~\ref{fig:rotational symmetry}a shows good agreement between the exact, rotationally symmetric $(X,Z)$ profiles and shapes found with nonlinear optimization of mesh-based energies across three levels of mesh refinement. 
Both the prolate (``peanut'') and oblate (``red blood cell'') shapes correspond to local minima of the Helfrich energy with $H_0=0$ and $k = 1$, subject to area and volume constraints quantified by reduced volume $v = 6\sqrt{\pi}V/A^{3/2}=0.75$ and length scale $R=(A/4\pi)^{1/2}=1$. 
To find minimal energy shapes with finer meshes, a (stretched, as in SI Sect.~\ref{sec:benchmark}) sphere with the desired number of vertices was interpolated to the minimum-energy shape found with a coarser mesh using the \texttt{interp} module.
Fig.~\ref{fig:rotational symmetry}b depicts the convergence of the minimal energy and the Lagrange multipliers found with numerical optimization to the exact values with increasing mesh refinement (number of vertices $n$). 
Results are nondimensionalized using the length scale $R$ and energy scale $E=2\pi k$ (Helfrich energy of a sphere).

We can also leverage the exact rotationally symmetric shapes to verify the force densities on the triangulated meshes found with autodiff. 
Fig.~\ref{fig:autodiff forces} compares the (signed) magnitude of the force densities (all forces act normal to the surface) arising from the Helfrich energy ($\mathbf{f}_H$), constant surface tension ($\mathbf{f}_A$), and constant pressure ($\mathbf{f}_V$) as calculated with the exact rotationally symmetric shape and autodiff on a triangulated mesh.
We see excellent agreement between the exact force densities and those found with autodiff. 
The strongest disagreement seen is between the pressure force density, which results from the rather simple expression $\mathbf{f}_V=p\hat{\mathbf{n}}$, inheriting the entire error of the Lagrange multiplier (It cannot be balanced by deviations in curvature like $\mathbf{f}_A$ and $\mathbf{f}_H$, which both involve curvature).
In each case, the force density is reported as a non-dimensional value $\mathbf{f}R^3/k$.
Furthermore, we observe the autodiff force densities sum to zero as expected for stationary shapes of the Helfrich energy (subject to volume and area constraints).

\subsection{Benchmark against an existing package for membrane mechanics simulation}
\label{sec:benchmark}

\begin{figure*}[tbh]
    \centering
    \includegraphics[width=\textwidth]{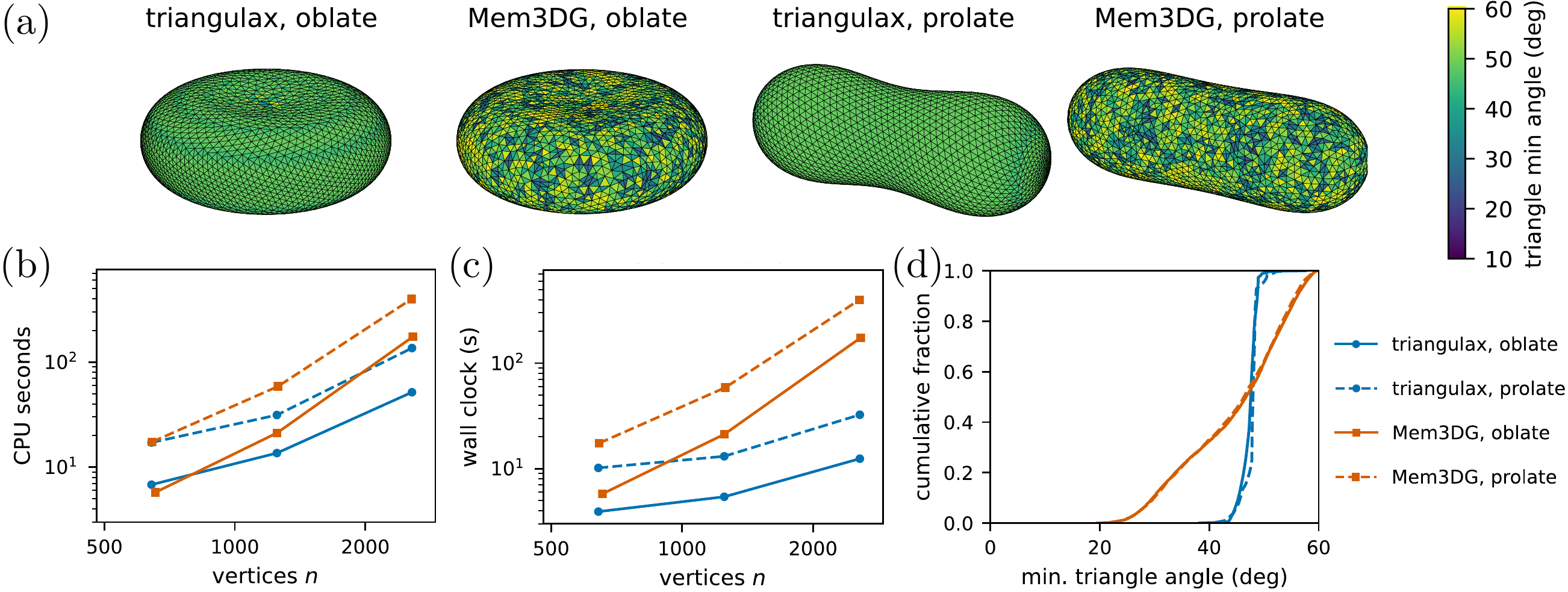}
    \caption{\textbf{Comparison of \tx and \memdg for membrane mechanics}.
    (a) Final equilibrium prolate and oblate shapes for \tx and \memdg. Each triangle is colored by its minimum angle on a shared linear scale. 
    (b--c) Process CPU seconds (b) and wall clock (c) for a repeat solve, which excludes \tx's one-off compilation. 
    (d) Cumulative probability distribution of the per-face minimum angle. 
    For (a) and (d), the number of vertices was $n=2562$ for \tx and comparable for \memdg.
    }
    \label{fig:benchmark supp}
\end{figure*}

We benchmark the performance of \tx against \memdg, an extensible framework for modeling 3d mechanochemical dynamics of membranes based on discrete differential geometry on triangulated meshes~\cite{Zhu.etal2022}.
\memdg is implemented in C++, which enables high computational efficiency but does not allow for automatic differentiation. 
To compare the performance of \tx and \memdg, we consider one of the most classic problems in membrane mechanics: deflation of a vesicle to prolate and oblate minimum-energy shapes (see Fig.~\ref{fig:membranes}(a) and Fig.~4 in ref.~\cite{Zhu.etal2022}). 
Specifically, the membrane is evolved by minimizing the Helfrich energy with volume and area penalties:
\begin{align}
    \label{eq:benchmark_energy}
    \begin{aligned}
    E = &\frac{k}{2} \int (H-H_0)^2 \, d{A} \\
      &+ \frac{K_A (A-A_0)^2}{2 A_0} 
      + K_V\left(\frac{V}{V_0} - \ln \frac{V}{V_0}- 1\right),
      \end{aligned}
\end{align}
where $A_0$ and $V_0$ are the target area and volume, and $K_A$ and $K_V$ are the magnitudes of area and volume penalties. $k$ is the bending stiffness and $H_0$ is the preferred curvature. 
Throughout this section, we take $A_0=4\pi$, $V_0 = \frac{4\pi}{3}\cdot 0.75$, $K_A=1$, $K_V=0.1$, $H_0=0$, and $k= 3.288 \times 10^{-4}$. 
All simulations are performed on a Mac Mini with Apple Silicon M4 Pro processors.

We initialize the simulations with a spherical Fibonacci lattice, which leads to a uniform mesh quality across the sphere~\cite{Swinbank2006_fibonaccimodelling}. 
To ensure convergence to prolate or oblate shapes, we perform a volume-preserving axisymmetric scaling $\mathrm{diag}(\rho^{-1/3}, \rho^{-1/3}, \rho^{2/3})$ and solve for $\rho$ so that the seed's discrete reduced volume ($v = 6\sqrt{\pi}\,V/A^{3/2}$) is exactly $v = 0.75$, taking $\rho > 1$ for the prolate branch and $\rho < 1$ for the oblate one.
To examine how performance scales with the mesh resolution, we repeat the calculation for different numbers of vertices: $n=642, 1250, 2562$.
Since the mesh mutator in \memdg does not preserve the number of vertices, we adjust the length scale for the mutator such that the final number of vertices is comparable to that of \tx. 
We set the mutation period to 500 integration steps. 

We first confirm that \tx and \memdg produce the same equilibrium shapes, with reduced volumes $v \approx 0.780$ for oblate shapes and  $v \approx 0.777$ for prolate shapes [Fig.~\ref{fig:benchmark supp}(a)]. 
Comparing the efficiency of the two packages  [Fig.~\ref{fig:benchmark supp}(b)--(c)], we find that the two packages are comparable at modest mesh resolutions, but \tx performs substantially better at large numbers of vertices ($n$).
Furthermore, $\tx$ also maintains a better mesh quality, with the minimal triangular angle concentrated between 40$^\circ$ and 50$^\circ$, while $\memdg$ exhibits a wider angle distribution and inhomogeneous mesh quality [Fig.~\ref{fig:benchmark supp}(d); see also Fig.~\ref{fig:benchmark supp}(a)].

\begin{figure*}[tbh]
    \centering
    \includegraphics[width=0.8\textwidth]{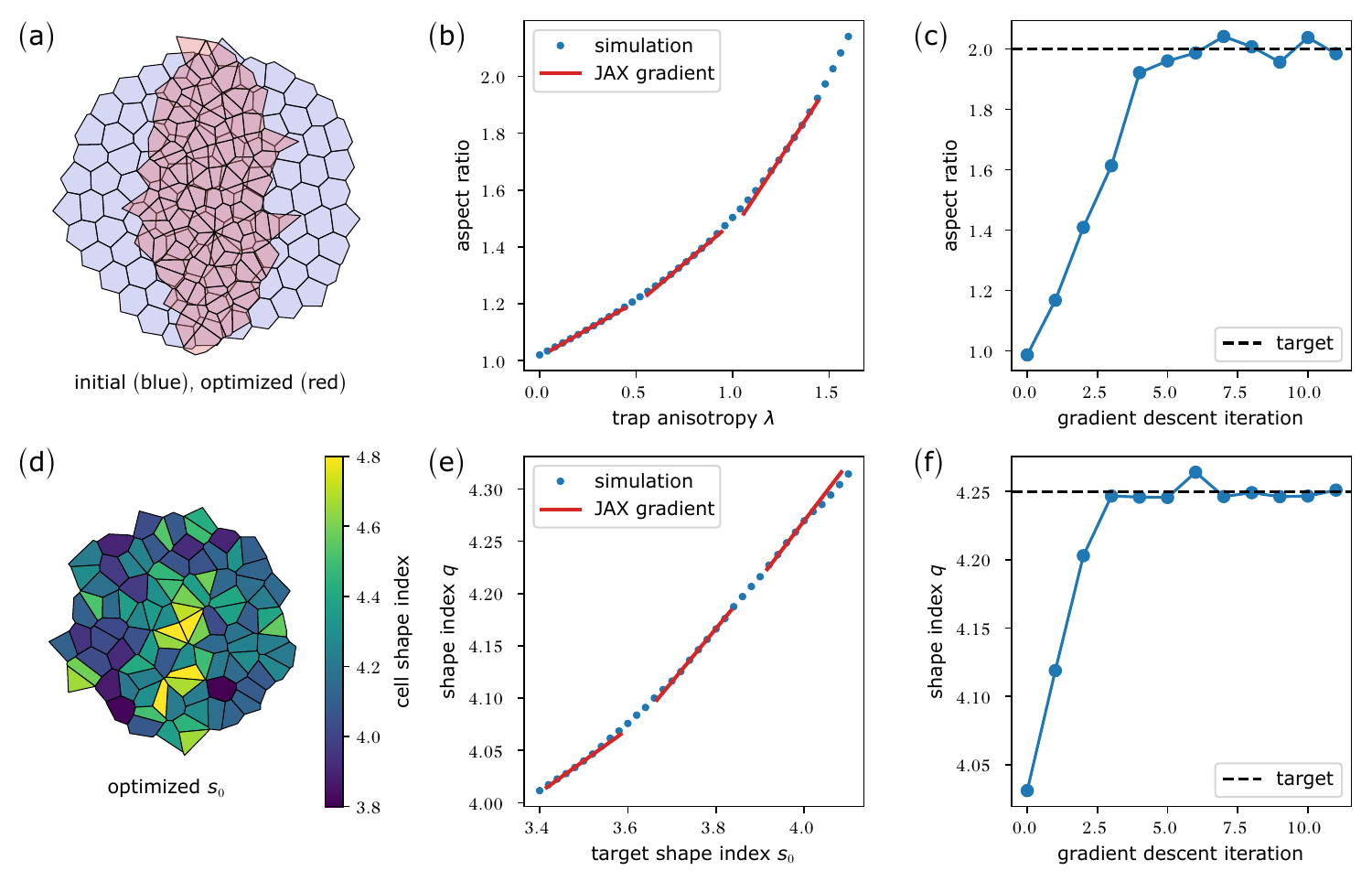}
    \caption{\textbf{Automatic differentiation of simulations with topological changes}.
    (a--c) show the inverse design of tissue aspect ratio, and (d--f) show the inverse design of shape parameter $s_0 = p_0/\sqrt{a_0}$. 
    These toy problems allow validating the simulation gradients computed by autodiff against finite differences and physical plausibility.
    (a)~Tissue shape for cells in an SVP model in a harmonic trap potential $V= \frac{k}{2}\left(e^{+\lambda}x^2 + e^{-\lambda} y^2\right)$ with anisotropy $\lambda$. 
    Blue cells show the final state of a simulation run with $\lambda=0$, including over 1,000 T1 transitions (edge flips) and stochastic self-propulsion forces. 
    We begin with an isotropic trap ($\lambda=0$), and then optimize $\lambda$ to design a target of aspect ratio $2$ in the simulation's final state ($\lambda\sim 1.5$, red cells). 
    (b)~\jx autodiff accurately computes the gradient of the tissue aspect ratio with respect to the anisotropic parameter $\lambda$ (red), in good agreement with that calculated by finite difference (blue).
    (c)~This allows optimized $\lambda$ to generate a desired aspect ratio of 2 in the final configuration by gradient descent. The simulation output for the optimized $\lambda$ is shown in red in (a).
    (d)~Optimizing the target shape parameter $s_0 = p_0/\sqrt{a_0}$ using autodiff, to bring the observed shape parameter $s$ (computed from cell shapes at simulation output) to $4.25$.
    (e)~Autodiff gradients match finite differences from simulations. In the fluid regime, the autodiff gradients become unreliable for very long simulations due to chaotic T1 dynamics. (Autodiff computes gradients for a ``frozen'' set of T1 transitions). Choosing shorter simulation times per gradient step avoids this issue.    
    (f)~Gradient-based optimization of model parameters to match the observed shape parameter to the target.
    }
    \label{fig:topo autodiff supp}
\end{figure*}

\end{document}